# DEGRADATION OF Ag/Si MULTILAYERS DURING HEAT TREATMENTS

K. Kapta, L. Daróczi, Z. Papp, D.L. Beke and G.A. Langer[a]

*Department of Solid State Physics, University of Debrecen, P.O. Box 2, H-4100 Debrecen, Hungary;*

A. Csik and M. Kis-Varga

*Institute of Nuclear Research of the Hungarian Academy of Sciences, P.O. Box. 51, H-4010 Debrecen, Hungary;*

A.L. Greer and Z.H. Barber

*Department of Materials Science and Metallurgy, University of Cambridge, Pembroke Street, Cambridge, CB2 3QZ, United Kingdom;*

PACS: 68.65 Ac; 42.70 Ln

**Abstract**

Microstructure changes during annealing of nano-crystalline silver and amorphous silicon multilayers (Ag/a-Si) have been studied by X-ray diffraction and transmission electron microscopy. The dc-magnetron sputtered Ag/a-Si multilayers remained stable even after annealing at 523K for 10h, and microstructural changes occurred only above 600K. The degradation of Ag/a-Si multilayers can be described by the increase of size of Ag grains, formation of grooves and pinholes at Ag grain boundaries and by the diffusion of silicon atoms through the silver grain boundaries and along the Ag/a-Si interfaces. This results in thinning of a-Si layers, and in formation of Ag granulates after longer annealing times.

---

[a] Corresponding author.
*E-mail address*: glanger@tigris.klte.hu

**Introduction**

The ability to create interfaces with well-defined structures is technologically important in synthesizing interfacial materials with unique optical, electronic, catalytic, magnetic and mechanical properties. On the other hand, the understanding of growth processes at surfaces is important for scientific reasons as well. In particular, many of the kinetic processes determining the thin-film structure are poorly understood. During recent years considerable experimental and theoretical efforts have been focused on investigation of metal/amorphous Si films that are miscible (W/Si, Mo/Si, Co/Si) [1,2,3]. But only a few reports about the metal/amorphous Si multilayers with practically no miscibility [4,5] have been published and our knowledge about their thermal stability or microstructural development is very little so far.

In the present work the Ag/ amorphous Si multilayer system, is considered. This structure is far from the thermodynamic equilibrium, since the ratio of the interface area to the volume is very large. Another contribution of the thermodynamic driving forces is given by the grain-boundary energy in fine-grained Ag layers. Besides these driving forces, the relief of mechanical (either growth or diffusion induced) stresses can influence the morphological development of multilayers. For this reason, during thermal annealing, microstructural changes are expected.

Zhao et al. [4.] have studied the thermal evolution of Ag/a-Si multilayers and found that the main diffusion mechanism was the diffusion of silicon atoms along the silver grain boundaries with a surprisingly small activation energy (0.24 eV) between 373 and 523 K, the silicon sublayers became thinner and silver sub-layers transformed into discontinuous ones.

In our experimental studies the microstructural development during thermal annealing of Ag/a-Si multilayers was measured by X-ray diffraction (XRD) and transmission electron microscopy (TEM).

**EXPERIMENTAL DETAILS**

Nano-crystalline Ag and amorphous Si films were deposited by alternate sputtering of Ag (99.95%) and Si (99.999%) in a dc magnetron system [6] onto NaCl and Si (100) substrates held at room temperature and located 70 mm from the targets. The base pressure was $3 \times 10^{-5}$ Pa and the Ar operating pressure was $7 \times 10^{-1}$ Pa. Multilayers with 12-25 Ag/Si layer pairs with thicknesses d(Ag)= 3-5 nm and d(Si)=3-5 nm were deposited on Si wafers. For the study of the agglomeration of Ag grains by means of plan-view transmission electron microscopy (TEM), a-Si/Ag/a-Si trilayers were sputtered onto NaCl substrates with single layer thickness of 4.5 nm. Plan-view transmission electron microscopy specimens were prepared by floating films off the NaCl substrate and picking them up on Cu TEM grids. The ion-beam milling for cross-sectional TEM pictures was carried out with liquid nitrogen cooling. The ion energy applied was less than 5 KeV. The annealing treatments were carried out in an atmosphere of high purity (99.999%) Ar and also in a $10^{-5}$ Pa vacuum. Various annealing treatments were performed at temperatures ranging from 423 to 773 K and with duration ranging from 30 min to 10 h. The temperature was measured by a NiCr-Ni thermocouple attached to the sample holder and controlled within a few K.

Microstructural characterization of the films was carried out using a JEOL 2000 transmission electron microscope equipped with an energy-dispersive X-ray spectrometer (EDS). X-ray diffraction instrument (XRD) with a Siemens Cu-anode X-ray tube was used to obtain XRD data in the θ-2θ geometry.

**Result and discussion**

Figure 1. shows the structure of the as-deposited Ag/a-Si multilayer together with selected area diffraction patterns (SADP). All the sharp rings present in the SADP correspond to the polycrystalline Ag layers. The halo corresponds to the amorphous Si. Figure 2a. represents the small-angle X-ray diffraction patterns (SAXRD). The sharp modulation peaks for the as grown samples indicate that a good quality multilayer has been produced. Figure 2b. shows the large-angle X-ray diffraction (LAXRD) scan of as-deposited Ag/a-Si as well as heat-treated multilayers indicating that the Ag crystallite size is about a few nm in all the samples. The development of Ag crystallite size during annealing is also illustrated in Fig. 2b. The average crystallite size, calculated from the width of the Ag X-ray peak by use of the Scherrer-formula $G_s=0.9\lambda/FWHM\cos\theta$, ($\lambda$ is the X ray wavelength, FWHM is the $2\theta$ full width at half maximum of the peak and $\theta$ is the Bragg angle), increases from a 3.5 nm to 6.5 nm.

The evolution of the SAXRD patterns in Figure 2a. shows that, up to 573 K the intensity and the ratio of small-angle Bragg peaks are practically unchanged; only the first order satellite increases slightly and the high order Bragg reflections become sharper. This means that phase separation of silver and silicon occurs, leading to the sharpening of the interfaces.

Above 600 K the second and fourth Bragg peaks start to decrease, indicating the change of the ratio of layer thickness of the silver and silicon.

The cross-sectional TEM images of the multilayer annealed at 723 K for 2 h show a peculiar change (Figure 3). It can be seen that in the upper part of the sample there is much less Ag, than in the bottom region just in contact with the substrate. (This was also confirmed by TEM +EDS). Furthermore the upper part of the sample is badly degraded. Although our results, obtained after annealing in high purity Ar and in vacuum, were the same the lack of silver is not clear at present and further study (e.g. improving the vacumm conditions) would be necessary to find the explanation. Due to the peculiar degradation in the upper part of the multilayer we discuss below processes taking part in the bottom part only.

At 723 K the first satellite decreases, the second one increases and the periodicity of the multilayer becomes worse. In this stage of the process the SAXRD patterns (in accordance with the TEM image in Figure 3.) practically reflect only the changes of the structure of that part of the sample which did not become degraded. During the annealing the position of the small-angle Bragg reflections is shifted towards larger angles, indicating the thinning of the multilayer period by about 6%. A similar effect was observed for miscible multilayers (W/a-Si, Mo/a-Si) in [7] and [8]. This was attributed to the presence of microcrystals in the amorphous Si layers. However, either by X-ray diffraction or by differential scanning calometry (Perkin Elmer DSC 7) measurements, we could not show the existence of microcrystalline Si. On the basis of the details of structural changes observed we emphasize that estimation of any diffusion coefficient from the decay law (the relation between the decay of intensity of first order Bragg peak and the interdiffusion coefficient [9]) is very questionable. This process is strictly reliable only for bulk intermixing, and thus it would lead to unrealistic results if applied to the degradation process observed by us. Thus the results of [4.] for the activation energies are very questionable.

Above 723 K the degradation process is very fast which suggests that the diffusional mass transport takes place along grain boundaries and interfaces and a granular structure, similar to the one shown in Fig. 4., develops. Furthermore, on the side of the Ag rich region pinholes can be observed (see the arrow in the inset of Figure 3.) which could be the initial stage of the granulation showed in Figure 4.

The large-angle X-ray diffraction patterns (Figure 2b.) indicate a gradual increase of the Ag crystallite size. Cross-sectional TEM observation shown the development of granular structure. These two facts indicate that although the formed granulates contain small crystallites as well, their diameter is larger than the initial crystallite size.

The results show that our samples are much more stable than those investigated by Zhao et. al. [4.], i.e. any changes of the multilayer structure occur at higher temperatures. For example, the formation of granular structure took place at temperatures about 200 K higher than reported in [4.]. It should be noted that, in their samples, the Ag/Si ratio was larger (as can be seen, in their TEM images) and their Ag/a-Si multilayers were prepared by ion beam sputtering deposition.

Plan-view TEM observations on a-Si4.5nm/Ag4.5nm/a-Si4.5nm trilayers showed that after heat treatment at 723 K for 1 h in high purity Ar the Ag layers agglomerate (Figure 4.). The separated Ag particles have diameters ranging from 10 to 80 nm (the mean diameter is about 50 nm). This degradation process was much faster than in the multilayer system for the same layer thickness. This fast granulation can be caused by the mechanical stresses (as-deposited or induced by the diffusion process) [10.].

**Conclusion**

Studies of morphological changes in a nanoscale Ag/a-Si layer system have shown that the as-deposited Ag/a-Si multilayers showed higher thermal stability than those produced in [4.], where the samples were produced by ion beam sputtering deposition and in their samples the Ag/Si ratio was larger than in our samples. In our case the microstructural development occurred only after heat treatment of the samples above 623 K/9 h, while in [4.] this process took place already at 373K.

Although we observed a peculiar degradation of the upper part of the multilayer, the following sequence could be established for the time evolution in the lower part (connected with the substrate):

i.)      sharpening of the interface, up to 573 K

ii.) increase of Ag crystallite size, diffusion of Si along Ag grain boundaries and pinhole formation, above 600 K

iii.) formation of the granular Ag layers by Si interface diffusion.

**Acknowledgment**

This work was supported by OTKA Grant No. T-025261, by the grant of the Hungarian Ministry of Education No. FKFP 0325/2000 and by Hungarian-British Intergovernmental and Cooperation Program No. GB-07/98. The authors thank G. Erdélyi for helpful discussions.

**REFERENCES**

1. Dupuis V, Ravet MF, Tete C, Piecuch M, Vidal BJ. Appl Phys 1990;68:3348-3355.
2. Holloway K, Ba Do K, Sinclair RJ. Appl Phys 1989;65:474-480.
3. Wang WH, Wang WKJ. Appl Phys 1994;76:1578.
4. Zhao JH, Zhang M, Liu RP, Zhang XY, Cao LM, Dai DY, Chen H, Xu YF, Wang WK. Journal of Materials Research-Pittsburgh 1999;14(7):2888-2892.
5. Wang WH, Bai HY, Wang WKJ. Appl Phys 1999;86:4262-4266.
6. Beke DL, Langer GA, Kis-Varga M, Dudas A, Nemes P, Daróczi L, Kerekes Gy, Erdélyi Z. Vacuum 1998;50:373-383.
7. Jiang Z, Jiang X, Liu W, Wu ZJ. Appl Phys 1989;65:196-200.
8. Kortright JB, Joksch St, Ziegler EJ. Appl Phys 1991;69:168-174.
9. Greer A, Spaepen F. Synthetic Modulated Structures. New York: Academic Press,1986.p.419.
10. Doener MF, Nix WD. Stresses and Deformation Processes in Thin Films on Substrates. Stanford University, 1986.

**FIGURE CAPTION**

Fig.1. Cross-sectional TEM image of an as-deposited Ag/a-Si multilayer (light layers are Si) and selected-area electron diffraction patterns.

Fig.2. Small- and large-angle XRD patterns of Ag/a-Si multilayer. a.) small-angle X-ray diffraction patterns, as deposited and heat treated (L modulation length), b.) large-angle X-ray diffraction pattern, as deposited and heat treated (Cs grain size calculated from the Scherrer-formula).

Fig.3. Cross-sectional TEM image of Ag/a-Si multilayer annealed at 723 K for 2 h. The inset shows the magnification of the bottom, Ag rich part of the annealed multilayer. The arrow indicates a pinhole.

Fig. 4. Plan-view TEM image of a-Si4.5nm/Ag4.5nm/a-Si4.5nm trilayer annealed at 723K for 1h.

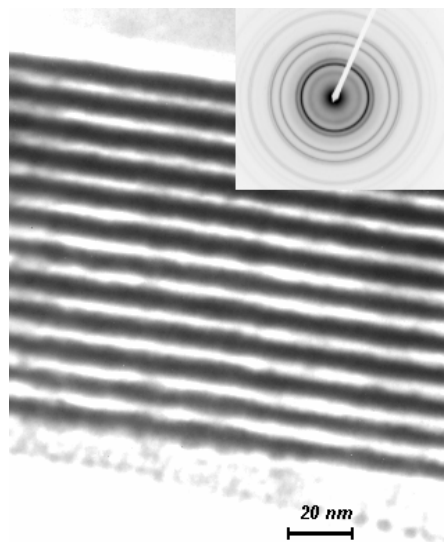

Fig.1. Cross-sectional TEM image of an as-deposited Ag/a-Si multilayer (light layers are Si) and selected-area electron diffraction patterns

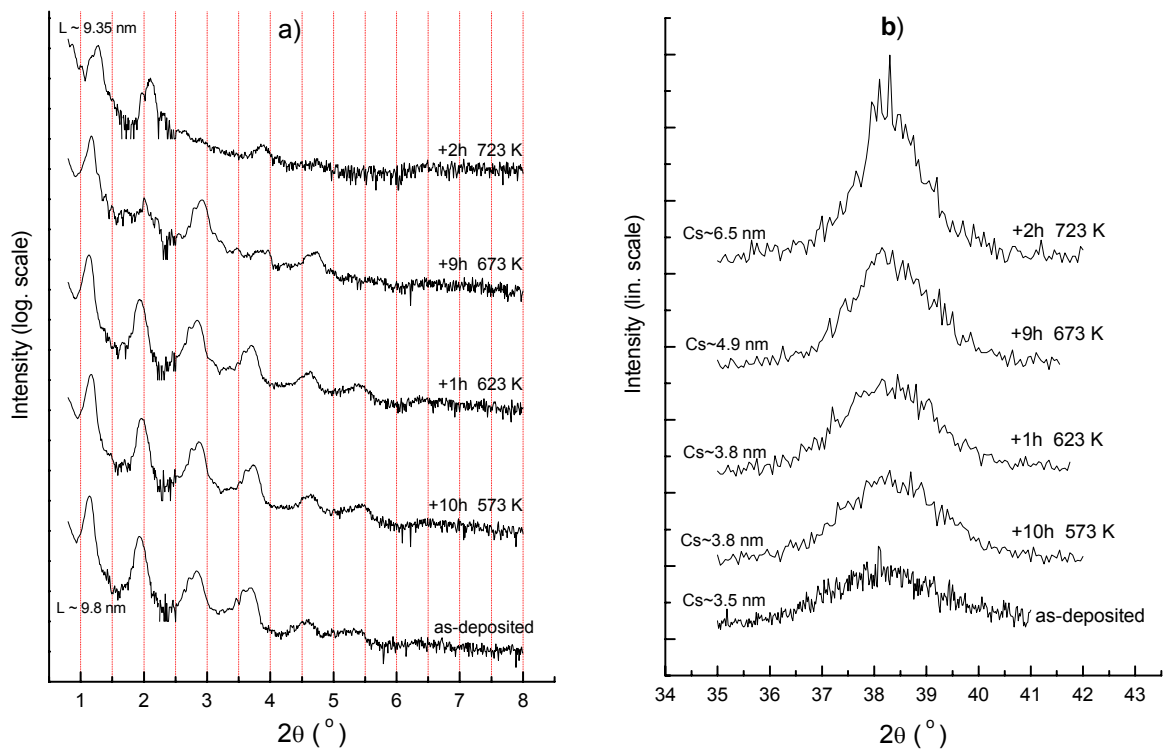

Fig.2. Small- and large-angle XRD patterns of Ag/a-Si multilayer. a.) small-angle X-ray diffraction patterns, as deposited and heat treated (L modulation length), b.) large-angle X-ray

diffraction pattern, as deposited and heat treated (Cs crystallite size calculated from Scherrer-formula).

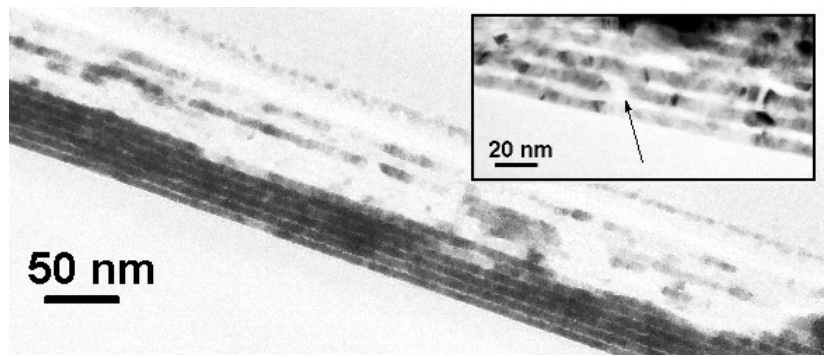

Fig. 3. Cross-sectional TEM image of Ag/a-Si multilayer annealed at 723 K for 2 h. The inset shows the magnification of the bottom, Ag rich part of the annealed multilayer. The arrow

indicates a pinhole.

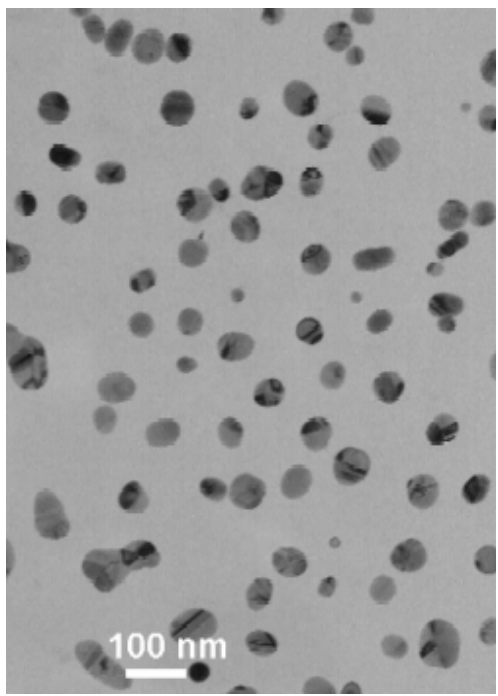

Fig.4. Plan-view TEM image of a-Si4.5nm/Ag4.5nm/a-Si4.5nm trilayer annealed at 723 K for 1h